\documentclass[amsmath,amssymb,floatfix,groupedaddress,twocolumn,prl]{revtex4}
\usepackage{graphicx}
\usepackage{color}

\begin{document}

\title{Accurate measurement of scattering and absorption loss in microphotonic devices}
\author{Matthew Borselli}
\thanks{Present address:  Xponent Photonics Inc., }
\author{Thomas J. Johnson}
\author{Oskar Painter}
\email{opainter@caltech.edu}
\affiliation{Department of Applied Physics, California Institute of Technology, Pasadena, CA 91125}

\date{\today}

\begin{abstract}  
We present a simple measurement and analysis technique to determine the fraction of optical loss due to both radiation (scattering) and linear absorption in microphotonic components.  The method is generally applicable to optical materials in which both nonlinear and linear absorption are present, and requires only limited knowledge of absolute optical power levels, material parameters, and the structure geometry.  The technique is applied to high quality factor ($Q=1$--$5 \times 10^6$) silicon-on-insulator microdisk resonators. It is determined that linear absorption can account for more than half the total optical loss in the high-$Q$ regime of these devices.
\end{abstract}

\maketitle

\noindent 
The push for dense integration of photonic elements into existing microelectronics circuits has revitalized the interest in semiconductor microphotonics\cite{Choi1,Xu1,Rong,ref:Fang1}.  Unfortunately, the benefits of tight optical confinement provided by high index contrast photonic elements have oftentimes been offset by increased optical losses due to high modal overlap with imperfect surfaces damaged by processing or imperfectly defined via lithography.  As advances in the etching and definition of these structures have reduced geometrical nonidealities, absorption has become a significant source of optical loss\cite{Borselli3}.  Understanding the optical losses of these structures is important for continued progress in developing low-loss microphotonic circuits; as a result, many recent articles\cite{Borselli2,Borselli3,Barrios,Alvarado,Michael1} have detailed methods for inferring the amount of optical loss due to absorption.  

Here we present a simple method for determining the linear absorption optical loss in microphotonic components without resorting to models of the thermal response of the structure or the character of the absorption.  The devices studied in this work consist of high-$Q$ silicon (Si) microdisk resonators formed from silicon-on-insulator (SOI) wafers.  As described in detail below, by monitoring the resonance wavelength and on-resonance transmission value of the microdisk modes as a function of input power, the relative amounts of linear absorption and radiation (scattering) loss within the resonator can be ascertained.  While this implementation of the method is specific to the silicon material system in which two-photon absorption (TPA)~\cite{ref:Dinu} is present, more generally it can be used in situations where there is some form of additional nonlinear absorption.  The technique is also applicable to a wide variety of resonator geometries, and thus can be employed for the study of optical loss in waveguides through the use of waveguide-based microring resonators.

\begin{figure}[ht]
\centerline{\includegraphics[width=8.3cm]{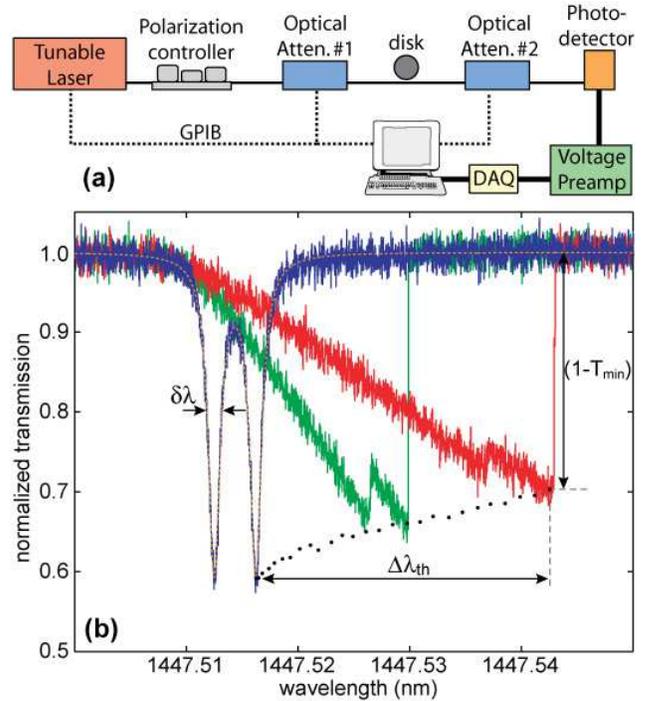}}
\caption{\label{fig:1}(a) Schematic representation of testing apparatus. (b) Measured transmission scans at various input powers for a 10-$\mu$m radius Si microdisk.  The input powers are $4$, $50$, and $100$ $\mu$W for the blue, green, and red curves, respectively.}
\end{figure}

The Si microdisks studied in this work consist of $10$-$\mu$m radius disks formed in a $217$-nm thick Si device layer on top of a $2$-$\mu$m SiO$_2$ BOX layer.  Details of the disk resonator fabrication process can be found in Refs. [\onlinecite{Borselli3}].  Device characterization was performed using a tunable external-cavity laser ($\lambda=$1420--1498 nm, linewidth $<$ 300 KHz for time scales relevant to this work) connected to a computer-controlled fiber taper waveguide probe\cite{Borselli2} and two optical attenuators as in Fig. \ref{fig:1}(a).  The micron-scale fiber taper probe was formed from a standard single-mode optical fiber and used to evanescently excite the whispering-gallery modes (WGMs) of the microdisk with controllable loading.  The two (highly linear) optical attenuators were controlled to provide variable optical input power to the resonators while maintaining a constant signal at the photodetector to eliminate nonlinearities in the detection electronics.

Figure \ref{fig:1}(b) shows the normalized spectral transmission response of a $10$-$\mu$m radius microdisk for several different input powers.  As is evident for the higher input powers, the thermo-optic effect of Si~\cite{Cocorullo} leads to an optical bistability in the transmission scan.  The observed doublet resonance dip, most discernible in the low power (blue) curve, is a result of surface roughness coupling the normally degenerate clockwise (CW) and counter-clockwise (CCW) propagating WGMs\cite{Gorodetsky}.  The resulting coupled modes can be conveniently described as sine- and cosine-like standing wave modes with their respective mode-field amplitudes ($a_s$ and $a_c$) given by

\begin{equation}
\label{eq:mode_amp}
a_{c(s)}=\frac{-\sqrt{\gamma_e/2}\sqrt{P_{i}}}{-(\gamma_t/2) + i(\Delta\omega\pm\gamma_{\beta}/2)},
\end{equation}

\noindent where $\Delta\omega$ is the detuning of the input laser frequency from the initially degenerate mode frequency $\omega_0$, $\gamma_t$ is the total decay rate of each individual resonance, $\gamma_e$ is the coupling rate into the fiber taper waveguide, $\gamma_{\beta}$ is the rate of CW to CCW mode coupling, $P_{i}$ is the optical input power at the fiber-cavity junction and we take the upper sign for $a_c$\cite{Borselli2}.  The normalized trasmission of the resonator-waveguide system is

\begin{equation}
\label{eq:Tran}
T=\frac{|-\sqrt{P_{i}}+\sqrt{\gamma_e/2}(a_c+a_s)|^2}{P_{i}}. 
\end{equation} 

\noindent Low power transmission data (blue curve in Fig. \ref{fig:1}(b)) is fit to eq. (\ref{eq:Tran}) and used to infer the resonator's total linear loss ($Q_t \equiv \omega_0/\gamma_t \approx \lambda_0/\delta\lambda=$1.5$\times10^6$) and the other parameters of eq. (\ref{eq:mode_amp}).  The resulting fit is displayed as the orange dashed curve in Fig. \ref{fig:1}(b).  For clarity of presentation, the analysis presented below is for singlet resonances: the analysis for doublet resonances is conceptually the same.

In order to facilitate the treatment below, we separate the total optical loss rate into a cold-cavity portion ($\gamma_c$) and a nonlinear absorption portion ($\gamma_{nla}$), $\gamma_t = \gamma_c + \gamma_{nla}$.  The cold-cavity loss rate comprises linear absorption ($\gamma_{la}$), radiation, and scattering loss.  Similarly, the absorption coefficient ($\gamma_a$) can be separated into linear and nonlinear contributions according to $\gamma_a=\gamma_{la}+\gamma_{nla}$.  With these definitions of loss rates we can write for the steady-state absorbed power within the cavity, $P_{\text{abs}}=(\gamma_a/\gamma_t)P_d$, where the dropped power ($P_d$) is related to the input power and normalized resonant transmission minimum ($T_{\text{min}}$) through $P_d=(1-T_{\text{min}})P_i$.

\begin{figure}
\centerline{\includegraphics[width=8.3cm]{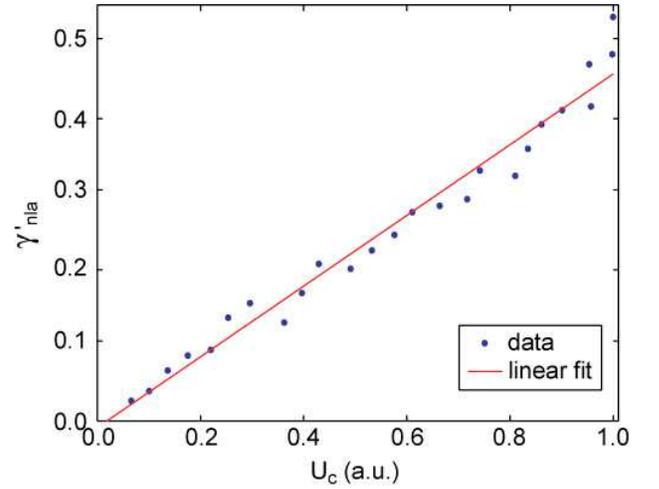}}
\caption{\label{fig:nonlin_abs} Plot of normalized nonlinear absorption versus relative electric-field cavity energy along with a linear fit.}
\end{figure}

Following Ref. \cite{Barclay6} we define the coupling factor, $K$, to be $(\gamma_e/\gamma_t)$.  Recalling our partitioning of the total loss rate, we find $\gamma'_{nla} = (K_0/K)-1$, where $\gamma'_{nla}$ is given by $(\gamma_{nla}/\gamma_{c})$ and $K_0$ is taken to be the coupling factor at low power in the absence of non-linear absorption, $K_0=(\gamma_e/\gamma_c)$.  Relating the coupling factor to the transmission minimum yields,

\begin{equation}
\label{eq:nonlin_to_tran}
\gamma'_{nla}(P_d) = K_0\left(\frac{1\mp\sqrt{(T_{\text{min}}(P_d))}}{1\pm\sqrt{(T_{\text{min}}(P_d))}}\right)-1,
\end{equation}

\noindent where the upper sign is taken in the undercoupled regime.  Thus we can infer the normalized non-linear absorption rate from the cold-cavity measurement of coupling factor and the observation of minimum transmission depth versus dropped power. Figure \ref{fig:nonlin_abs} shows a plot of the normalized nonlinear absorption of the 10-$\mu$m Si microdisk of Fig. \ref{fig:1} measured using this technique.  It should be noted that the absolute dropped power need not be known to find this quantity.  The relative dropped power can be converted into a relative cavity energy ($U_{c}=P_d/\gamma_t$) to observe the energy dependence of the normalized nonlinear absorption, as is done in Fig. \ref{fig:nonlin_abs}.  The linear dependence of $\gamma'_{nla}$ versus $U_{c}$ indicates that TPA is the dominant nonlinear absorption inside the Si microdisks at these input powers (peak input power of $\sim 100$ $\mu$W).

The remaining quantity of interest, $\gamma_a$, can be determined by considering the relationship between the resonance position and absorbed power due to the thermo-optic effect of Si.  For small temperature changes we expect the resonance wavelength shift, $\Delta\lambda$, to be linearly proportional to the absorbed power~\cite{Cocorullo}. Therefore, we have $\Delta\lambda=C(\gamma_a/\gamma_t)P_{d}$, where $C$ is a constant depending upon the thermal and thermo-optic characteristics of the microdisks.  Keeping in mind the decomposition of $\gamma_a$ and $\gamma_t$ above, we find 

\begin{equation}
\label{eq:wave_shift}
\Delta\lambda(P_d)=C\left(\frac{\gamma'_{la} + \gamma'_{nla}(P_d)}{1+\gamma'_{nla}(P_d)}\right)P_d,
\end{equation}

\noindent where $\gamma'_{la}$ is given by $({\gamma_{la}/\gamma_c})$.  With $\gamma'_{nla}(P_d)$ already measured from the resonant transmission versus input power, eq. (\ref{eq:wave_shift}) indicates that $\gamma'_a$ may be determined directly from the global slope curve, $[\Delta\lambda_{th} / P_d](P_d)$.  As $\gamma'_a$ can be estimated from this slope curve alone, we need not know the absolute dropped power.  Similarly, knowledge of the temperature change of the resonator and the thermal resistance of the structure, needed for directly estimating the absorbed power, is also not needed. 

\begin{figure}[t]
\centerline{\includegraphics[width=8.3cm]{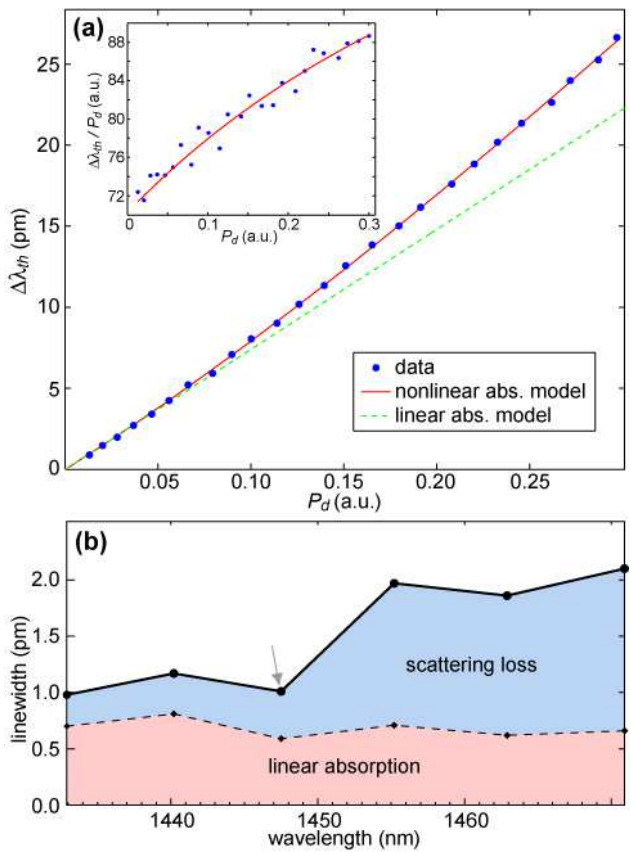}}
\caption{\label{fig:thermal_shift} (a) Plot of thermally-induced wavelength shift ($\Delta\lambda_{th}$) versus relative dropped power ($P_d$) along with nonlinear and linear absorption model fits. (inset) Global slope, $\Delta\lambda_{th} / P_d$, versus $P_d$ for the same dataset. (b) Measured intrinsic linewidth for the entire family of high-$Q$ WGMs of this microdisk, along with the measured delineation between scattering loss and linear absorption.  The resonant mode of panel (a) is denoted with a grey arrow in this plot.}
\end{figure}

Figure \ref{fig:thermal_shift}(a) shows the measured resonance shift from Fig. \ref{fig:1}(b) versus $P_d$ along with a fit to eq. (\ref{eq:wave_shift}) and a linear fit to the lowest power data as a visual aid.  The inset also plots the global slope curve, $[\Delta\lambda_{th} / P_d](P_d)$, for the same dataset.  From a fit to the global slope curve we find that $\gamma'_{la}=0.57\pm0.03$ for this resonant mode.  Analysis of the entire family of WGMs (varying azimuthal mode number, fixed radial order and polarization) for the same microdisk is plotted in Fig. \ref{fig:thermal_shift}(b).  Linear absorption is found to persist at a nearly constant level of $\delta\lambda_{a}=0.5$ pm across the 1400-nm wavelength band, corresponding to a loss per unit length of roughly $0.25$ dB/cm.  As bulk absorption in the moderately p-doped Si of these devices is expected to be more than an order of magnitude smaller~\cite{Soref2} than this measured value, the linear absorption loss is attributed to the etched and polished surfaces of the disk.  This is consistent with recent studies we performed on similar devices in which optical absorption loss could be dramatically varied with changing chemistry of the Si surfaces~\cite{Borselli3}.  

As the above example serves to illustrate, for materials in which there is a measureable amount of two-photon absorption, by monitoring the resonance wavelength and transmission of a microresonator one can separate the amount of linear absorption from that of other loss components such as intrinsic radiation and scattering.  More generally, this technique can provide insights into the nature of optical loss in other materials and devices utilizing \textit{nonlinear} absorption to accurately determine \textit{linear} absorption with only limited knowledge of absolute power, material parameters, and physical geometry of the structure.

This work was supported by DARPA through the EPIC program. The authors thank Chris Michael, Paul Barclay and Kartik Srinivasan for useful discussions.  M.B. thanks the Moore Foundation, NPSC, and HRL Laboratories for his graduate fellowship support.


\begin{thebibliography}{14}
\expandafter\ifx\csname natexlab\endcsname\relax\def\natexlab#1{#1}\fi
\expandafter\ifx\csname bibnamefont\endcsname\relax
  \def\bibnamefont#1{#1}\fi
\expandafter\ifx\csname bibfnamefont\endcsname\relax
  \def\bibfnamefont#1{#1}\fi
\expandafter\ifx\csname citenamefont\endcsname\relax
  \def\citenamefont#1{#1}\fi
\expandafter\ifx\csname url\endcsname\relax
  \def\url#1{\texttt{#1}}\fi
\expandafter\ifx\csname urlprefix\endcsname\relax\def\urlprefix{URL }\fi
\providecommand{\bibinfo}[2]{#2}
\providecommand{\eprint}[2][]{\url{#2}}

\bibitem[{\citenamefont{Xu et~al.}(2005)\citenamefont{Xu, Schmidt, Pradhan, and
  Lipson}}]{Xu1}
\bibinfo{author}{\bibfnamefont{Q.}~\bibnamefont{Xu}},
  \bibinfo{author}{\bibfnamefont{B.}~\bibnamefont{Schmidt}},
  \bibinfo{author}{\bibfnamefont{S.}~\bibnamefont{Pradhan}}, \bibnamefont{and}
  \bibinfo{author}{\bibfnamefont{M.}~\bibnamefont{Lipson}},
  \bibinfo{journal}{Nature} \textbf{\bibinfo{volume}{435}},
  \bibinfo{pages}{325} (\bibinfo{year}{2005}).

\bibitem[{\citenamefont{Rong et~al.}(2005)\citenamefont{Rong, Liu, Jones,
  Cohen, Hak, Nicolaescu, Fang, and Paniccia}}]{Rong}
\bibinfo{author}{\bibfnamefont{H.}~\bibnamefont{Rong}},
  \bibinfo{author}{\bibfnamefont{A.}~\bibnamefont{Liu}},
  \bibinfo{author}{\bibfnamefont{R.}~\bibnamefont{Jones}},
  \bibinfo{author}{\bibfnamefont{O.}~\bibnamefont{Cohen}},
  \bibinfo{author}{\bibfnamefont{D.}~\bibnamefont{Hak}},
  \bibinfo{author}{\bibfnamefont{R.}~\bibnamefont{Nicolaescu}},
  \bibinfo{author}{\bibfnamefont{A.}~\bibnamefont{Fang}}, \bibnamefont{and}
  \bibinfo{author}{\bibfnamefont{M.}~\bibnamefont{Paniccia}},
  \bibinfo{journal}{Nature} \textbf{\bibinfo{volume}{433}},
  \bibinfo{pages}{292} (\bibinfo{year}{2005}).

\bibitem[{\citenamefont{Choi et~al.}(2004)\citenamefont{Choi, Djrodjev, Peng,
  Yang, Choi, and Dapkus}}]{Choi1}
\bibinfo{author}{\bibfnamefont{S.~J.} \bibnamefont{Choi}},
  \bibinfo{author}{\bibfnamefont{K.~D.} \bibnamefont{Djrodjev}},
  \bibinfo{author}{\bibfnamefont{Z.}~\bibnamefont{Peng}},
  \bibinfo{author}{\bibfnamefont{Q.}~\bibnamefont{Yang}},
  \bibinfo{author}{\bibfnamefont{S.~J.} \bibnamefont{Choi}}, \bibnamefont{and}
  \bibinfo{author}{\bibfnamefont{P.~D.} \bibnamefont{Dapkus}},
  \bibinfo{journal}{IEEE Photonics Tech. Lett.} \textbf{\bibinfo{volume}{17}},
  \bibinfo{pages}{2101} (\bibinfo{year}{2004}).

\bibitem[{\citenamefont{Fang et~al.}(2006)\citenamefont{Fang, Park, Cohen,
  Jones, Paniccia, and Bowers}}]{ref:Fang1}
\bibinfo{author}{\bibfnamefont{A.~W.} \bibnamefont{Fang}},
  \bibinfo{author}{\bibfnamefont{H.}~\bibnamefont{Park}},
  \bibinfo{author}{\bibfnamefont{O.}~\bibnamefont{Cohen}},
  \bibinfo{author}{\bibfnamefont{R.}~\bibnamefont{Jones}},
  \bibinfo{author}{\bibfnamefont{M.~J.} \bibnamefont{Paniccia}},
  \bibnamefont{and} \bibinfo{author}{\bibfnamefont{J.~E.}
  \bibnamefont{Bowers}}, \bibinfo{journal}{Opt. Express}
  \textbf{\bibinfo{volume}{14}}, \bibinfo{pages}{9203} (\bibinfo{year}{2006}).

\bibitem[{\citenamefont{Borselli et~al.}(2006)\citenamefont{Borselli, Johnson,
  and Painter}}]{Borselli3}
\bibinfo{author}{\bibfnamefont{M.}~\bibnamefont{Borselli}},
  \bibinfo{author}{\bibfnamefont{T.~J.} \bibnamefont{Johnson}},
  \bibnamefont{and} \bibinfo{author}{\bibfnamefont{O.}~\bibnamefont{Painter}},
  \bibinfo{journal}{Appl. Phys. Lett.} \textbf{\bibinfo{volume}{88}},
  \bibinfo{pages}{131114} (\bibinfo{year}{2006}).

\bibitem[{\citenamefont{Borselli et~al.}(2005)\citenamefont{Borselli, Johnson,
  and Painter}}]{Borselli2}
\bibinfo{author}{\bibfnamefont{M.}~\bibnamefont{Borselli}},
  \bibinfo{author}{\bibfnamefont{T.~J.} \bibnamefont{Johnson}},
  \bibnamefont{and} \bibinfo{author}{\bibfnamefont{O.}~\bibnamefont{Painter}},
  \bibinfo{journal}{Opt. Express} \textbf{\bibinfo{volume}{13}},
  \bibinfo{pages}{1515} (\bibinfo{year}{2005}).

\bibitem[{\citenamefont{Barrios et~al.}(2004)\citenamefont{Barrios, Almeida,
  Panepucci, Schmidt, and Lipson}}]{Barrios}
\bibinfo{author}{\bibfnamefont{C.~A.} \bibnamefont{Barrios}},
  \bibinfo{author}{\bibfnamefont{V.~R.} \bibnamefont{Almeida}},
  \bibinfo{author}{\bibfnamefont{R.~R.} \bibnamefont{Panepucci}},
  \bibinfo{author}{\bibfnamefont{B.~S.} \bibnamefont{Schmidt}},
  \bibnamefont{and} \bibinfo{author}{\bibfnamefont{M.}~\bibnamefont{Lipson}},
  \bibinfo{journal}{IEEE Photonics Tech. Lett.} \textbf{\bibinfo{volume}{16}},
  \bibinfo{pages}{506} (\bibinfo{year}{2004}).

\bibitem[{\citenamefont{Alvarado-Rodriguez and Yablonovitch}(2002)}]{Alvarado}
\bibinfo{author}{\bibfnamefont{I.}~\bibnamefont{Alvarado-Rodriguez}}
  \bibnamefont{and}
  \bibinfo{author}{\bibfnamefont{E.}~\bibnamefont{Yablonovitch}},
  \bibinfo{journal}{J. Appl. Phys.} \textbf{\bibinfo{volume}{92}},
  \bibinfo{pages}{6399} (\bibinfo{year}{2002}).

\bibitem[{\citenamefont{Michael et~al.}(2007)\citenamefont{Michael, Srinivasan,
  Johnson, Painter, Lee, Hennessy, Kim, and Hu}}]{Michael1}
\bibinfo{author}{\bibfnamefont{C.~P.} \bibnamefont{Michael}},
  \bibinfo{author}{\bibfnamefont{K.}~\bibnamefont{Srinivasan}},
  \bibinfo{author}{\bibfnamefont{T.~J.} \bibnamefont{Johnson}},
  \bibinfo{author}{\bibfnamefont{O.}~\bibnamefont{Painter}},
  \bibinfo{author}{\bibfnamefont{K.~H.} \bibnamefont{Lee}},
  \bibinfo{author}{\bibfnamefont{K.}~\bibnamefont{Hennessy}},
  \bibinfo{author}{\bibfnamefont{H.}~\bibnamefont{Kim}}, \bibnamefont{and}
  \bibinfo{author}{\bibfnamefont{E.}~\bibnamefont{Hu}}, \bibinfo{journal}{Appl.
  Phys. Lett.} \textbf{\bibinfo{volume}{90}} (\bibinfo{year}{2007}).

\bibitem[{\citenamefont{Dinu et~al.}(2003)\citenamefont{Dinu, Quochi, and
  Garcia}}]{ref:Dinu}
\bibinfo{author}{\bibfnamefont{M.}~\bibnamefont{Dinu}},
  \bibinfo{author}{\bibfnamefont{F.}~\bibnamefont{Quochi}}, \bibnamefont{and}
  \bibinfo{author}{\bibfnamefont{H.}~\bibnamefont{Garcia}},
  \bibinfo{journal}{Appl. Phys. Lett.} \textbf{\bibinfo{volume}{82}},
  \bibinfo{pages}{2954} (\bibinfo{year}{2003}).

\bibitem[{\citenamefont{Cocorullo and Rendina}(1992)}]{Cocorullo}
\bibinfo{author}{\bibfnamefont{G.}~\bibnamefont{Cocorullo}} \bibnamefont{and}
  \bibinfo{author}{\bibfnamefont{I.}~\bibnamefont{Rendina}},
  \bibinfo{journal}{IEE Elec. Lett.} \textbf{\bibinfo{volume}{28}},
  \bibinfo{pages}{83} (\bibinfo{year}{1992}).

\bibitem[{\citenamefont{Gorodetsky et~al.}(2000)\citenamefont{Gorodetsky,
  Pryamikov, and Ilchenko}}]{Gorodetsky}
\bibinfo{author}{\bibfnamefont{M.}~\bibnamefont{Gorodetsky}},
  \bibinfo{author}{\bibfnamefont{A.}~\bibnamefont{Pryamikov}},
  \bibnamefont{and} \bibinfo{author}{\bibfnamefont{V.}~\bibnamefont{Ilchenko}},
  \bibinfo{journal}{J. Opt. Soc. Am. B} \textbf{\bibinfo{volume}{17}},
  \bibinfo{pages}{1051} (\bibinfo{year}{2000}).

\bibitem[{\citenamefont{Barclay et~al.}(2005)\citenamefont{Barclay, Srinivasan,
  and Painter}}]{Barclay6}
\bibinfo{author}{\bibfnamefont{P.~E.} \bibnamefont{Barclay}},
  \bibinfo{author}{\bibfnamefont{K.}~\bibnamefont{Srinivasan}},
  \bibnamefont{and} \bibinfo{author}{\bibfnamefont{O.}~\bibnamefont{Painter}},
  \bibinfo{journal}{Opt. Express} \textbf{\bibinfo{volume}{13}},
  \bibinfo{pages}{801} (\bibinfo{year}{2005}).

\bibitem[{\citenamefont{Soref and Bennett}(1987)}]{Soref2}
\bibinfo{author}{\bibfnamefont{R.~A.} \bibnamefont{Soref}} \bibnamefont{and}
  \bibinfo{author}{\bibfnamefont{B.~R.} \bibnamefont{Bennett}},
  \bibinfo{journal}{IEEE J. Quan. Elec.} \textbf{\bibinfo{volume}{23}},
  \bibinfo{pages}{123} (\bibinfo{year}{1987}).

\end{thebibliography}
\end{document}